\documentclass[twocolumn,showpacs,amsmath,amssymb,aps,prl]{revtex4}
\usepackage{graphicx}
\usepackage{dcolumn}
\usepackage{bm}
\usepackage{wrapfig,subfigure}
\usepackage{amssymb}
\usepackage{color}

\newcommand{\n}{\mbox{\boldmath $\nabla$}}
\newcommand{\qb}{{\bf q}}
\newcommand{\pb}{{\bf p}}
\newcommand{\fb}{{\bf f}}

\newcommand{\xib}{\bm\xi}

\def\p12{p_{12}({\bf q},t)}

\begin{document}

\title{Paths of fluctuation induced switching}

\author{H. B. Chan$^{(1)}$, M. I. Dykman$^{(2)}$, and C. Stambaugh$^{(1)}$}
\affiliation{$^{(1)}$Department of Physics, University of Florida, Gainesville, FL 32611\\
$^{(2)}$Department of Physics and Astronomy, Michigan State University, East Lansing, MI 48823}

\begin{abstract}
We demonstrate that the paths followed by a system in fluctuation-activated switching form a narrow tube in phase space. A theory of the path distribution is developed and its direct measurement is performed in a micromechanical oscillator. The experimental and theoretical results are in excellent agreement, with no adjustable parameters. We also demonstrate the lack of time-reversal symmetry in switching of systems far from thermal equilibrium.
\end{abstract}

\maketitle

Fluctuation-activated switching between coexisting stable states is at the root of diverse phenomena, from switching in nanomagnets \cite{Wernsdorfer1997} and Josephson junctions \cite{Fulton1974} to chemical reactions and to protein folding \cite{Wales2003}. A detailed theory of switching rates was first developed by Kramers \cite{Kramers1940} for systems close to thermal equilibrium. Here, the switching rate is determined by the free energy barrier between the states. In recent years much attention has been given to switching in systems far from thermal equilibrium, like electrons \cite{Lapidus1999} and atoms \cite{Gommers2005,Kim2006} in modulated traps and rf-driven Josephson junctions \cite{Siddiqi2004,Lupascu2007} and nano- and micromechanical resonators \cite{Aldridge2005,Stambaugh2006,Almog2007}. Nonequilibrium systems generally lack detailed balance, and the switching rates may not be found by a simple extension of the Kramers approach.

The analysis of switching in both nonequilibrium systems and complex equilibrium systems, including biomolecules, relies on the idea that, even though the system motion is random, the switching paths form narrow tubes in phase space. The tube is centered at the most probable switching path (MPSP). For low fluctuation intensity, the MPSP is obtained from a variational problem, which also determines the switching activation barrier \cite{Freidlin_book,Dykman1979a,Graham1984a,Bray1989,Maier1993a,Kraut2004,Dykman2004,Bier2005,Tretiakov2005,Elgart2006}. Despite its fundamental role, the concept of the narrow tube of switching paths has not been tested experimentally nor has this tube been characterized quantitatively. Prior experimental studies \cite{Hales2000} and simulations \cite{Luchinsky1997b,Morillo1997} focused on the distribution in space and time of fluctuational paths to a certain space-time point \cite{Dykman1992d}. The methods \cite{Hales2000,Luchinsky1997b,Morillo1997,Dykman1992d} do not apply to the switching paths distribution for systems with more than one dynamical variable because of the motion slowing down near the stable states and the related loss of path synchronization.

\begin{figure}%
\includegraphics[width=3.2in]{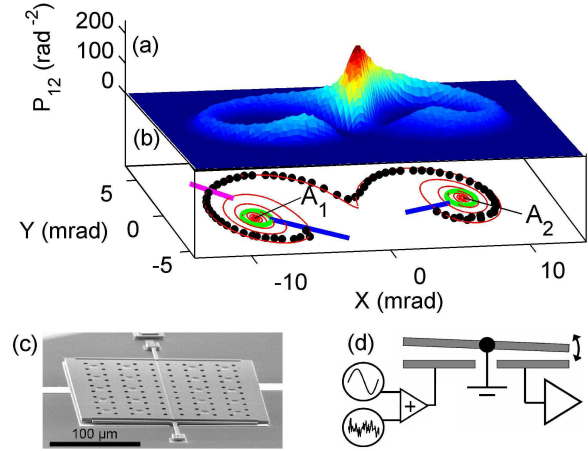}
\caption{\label{fig:1} (a) Switching probability distribution in a parametrically driven microelectromechanical oscillator. The distribution  $p_{12}(X,Y)$ is measured for switching out of state $A_1$  into state  $A_2$. (b) The peak locations of the distribution are plotted as black circles and the theoretical most probable switching path is indicated by the red line. All trajectories originate from within the green circle in the vicinity of  $A_1$ and later arrive at the green circle around  $A_2$. The radii of the green circles give the typical fluctuation amplitude $l_D$. The portion of the distribution outside the blue lines is omitted. (c) Scanning electron micrograph of the micromechanical torsional oscillator. (d) Cross-sectional schematic.}
\end{figure}

In this paper, we present a theory of the switching paths distribution and report its direct observation in a well-characterized system, a micromechanical oscillator driven into parametric resonance. The observed distribution is shown in Fig.~\ref{fig:1}a. Our experimental and theoretical results are in excellent agreement, with no adjustable parameters. In addition to proving the concept of the narrow tube of switching paths and putting it on a quantitative basis, the results provide the first demonstration of the lack of time-reversal symmetry in switching of systems far from thermal equilibrium. The results also open the possibility of efficient control of the switching probability based on the measured narrow path distribution.

We consider a bistable system with several dynamical variables  $\qb = (q_1,...,q_N)$. The stable states $A_1$  and $A_2$ are located at $\qb_{A_1}$  and $\qb_{A_2}$ respectively. For low fluctuation intensity, the physical picture of switching is as follows. The system prepared initially near state  $A_1$, for example, would fluctuate about it with small typical amplitude $l_D$. Ultimately, over time of order of the reciprocal switching rate $W_{12}^{-1}$   that largely exceeds the relaxation time  $t_r$, the system will make a transition to state $A_2$  and will then fluctuate about it. In the transition the system first moves from the vicinity of $\qb_{A_1}$  to the vicinity of the saddle point $\qb_{S}$  as a result of a large and rare fluctuation. Its trajectory is expected to be close to the one for which the probability of the appropriate fluctuation is maximal. From the vicinity of $\qb_{S}$  the system moves to state $A_2$  close to the deterministic fluctuation-free ``downhill'' trajectory. These two trajectories comprise the MPSP.

We characterize the switching paths distribution by the conditional probability density for the system to pass through a point $\qb$  on its way from $A_1$  to  $A_2$,
\begin{equation}
\label{eq:1}
 p_{12}(\qb,t)=\int_{\Omega_2}d\qb_{f}\rho(\qb_{f},t_f;\qb,t\vert \qb_{0},t_0).
\end{equation}
Here, the integrand is the probability density for the system to be at points  $\qb_{f}$  and  $\qb$  at times $t_f$  and  $t$, respectively, given that it was at   $\qb_{0}$ at time  $t_0$. The point   $\qb_{0}$ lies within distance $\sim l_D$  of  $\qb_{A_1}$  and is otherwise arbitrary. Integration with respect to  $\qb_{f}$  goes over the range  $\Omega_2$ of small fluctuations about   $\qb_{A_2}$.

For  $W_{12}^{-1},W_{21}^{-1} \gg t_f-t,t-t_0 \gg t_r$, far from the stable states the distribution (\ref{eq:1}) is independent of  $\qb_0$, $t_0,\, t_f$, and  $t$. It peaks on the MPSP. The peak is Gaussian transverse to the MPSP. For  $|\qb-\qb_{A_{1,2}}|, |\qb-\qb_{S}|\gg l_D$
\begin{equation}
\label{eq:2}
 p_{12}(\qb,t)= W_{12}v^{-1}(\xi_\|)Z^{-1}\exp\left({-\frac{1}{2}
\xib_{\perp}\widehat{Q}\xib_{\perp}}\right),
\end{equation}
where $\xi_{\|}$  and  $\xib_{\perp}$  are coordinates along and transverse to the MPSP, and $v(\xi_\|)$  is the velocity along the MPSP. The matrix elements of matrix  $\widehat{Q}=\widehat{Q}(\xi_\|)$ are  $\propto l_D^{-2}$, and  $Z=[(2\pi)^{N-1}/\rm{det}\widehat{Q}]^{1/2}$.

We derive Eq.~(\ref{eq:2}) for a system described by the Langevin equation of motion %
\begin{eqnarray}
\label{eq:Langevin}
 \dot \qb = {\bf K}(\qb) + \fb(t), \; \langle f_n(t) f_m(t')\rangle = 2D\delta_{nm}\delta(t-t').
\end{eqnarray}
Here, $\fb(t)$ is white Gaussian noise. The noise intensity $D$ is assumed small. The dependence of the switching rates $W_{ij}$ on $D$ is given by the activation law, $\log W_{ij}\propto D^{-1}$ \cite{Freidlin_book}. The length $l_D=(Dt_r)^{1/2}$ where the relaxation time $t_r$ is given by the reciprocal minimal (in absolute value) real part of the eigenvalues of the matrix $\partial K_m/\partial q_n$ calculated at $\qb_{A_{1,2}}, \qb_S$ where ${\bf K}(\qb)={\bf 0}$.

The integrand in Eq.~(1) can be written as a product of transition probabilities,
$ \rho(\qb_f,t_f;\qb,t|\qb_0,t_0)=\rho(\qb_f,t_f|\qb,t)
\rho(\qb,t|\qb_0,t_0)$.
In the time range of interest they satisfy the stationary backward and forward Fokker-Planck equations, respectively,
\begin{eqnarray}
\label{eq:FPE}
 &&\left[{\bf K}\partial_{\qb} +
 D\partial^2_{\qb}\right]\rho(\qb_f,t_f|\qb,t)=0,\nonumber\\  &&[-\partial_{\qb}{\bf K}+D\partial_{\qb}^2]\rho(\qb,t|\qb_0,t_0) =  0.
\end{eqnarray}

We start with the case where the observation point $\qb$ lies in the region of attraction of the initially empty state $A_2$. Here, $\rho(\qb_f,t_f|\qb,t) \approx \rho_2(\qb_f)$, where $\rho_2(\qb_f)$ is the normalized quasistationary probability distribution in the state $A_2$ in the neglect of interstate switching. Then, from Eq.~(1) $\p12 =\rho(\qb,t|\qb_0,t_0)$.

The function $ \rho(\qb,t|\qb_0,t_0)$ describes a stationary probability current from the vicinity of $\qb_S$ to the attractor $\qb_{A_2}$. It is concentrated on the noise-free ``downhill" trajectory $\dot\qb = {\bf K}$ coming out of the saddle, which thus gives the corresponding section of the MPSP. Diffusion leads to broadening of the distribution $ \rho(\qb,t|\qb_0,t_0)$ \cite{Ludwig1975}. The total current gives the switching rate $W_{12}$ \cite{Kramers1940}.

On the downhill trajectory, $\hat{\bm\xi}_{\parallel}$ points along the vector ${\bf K}$ and the velocity is $v(\xi_{\parallel})= K(\xi_{\parallel},{\bm\xi}_{\perp}={\bf 0})$. For small $\xi_{\perp}\lesssim l_D$ the solution of Eq.~(\ref{eq:FPE}) has the form (2), with matrix $\widehat Q$ given by equation
$ v\,\partial_{\xi_{\parallel}}\widehat{Q}+\hat \kappa^{\dagger}\widehat{Q} + \widehat{Q}\hat\kappa + 2\widehat{Q}^2D=0 $,
where $\kappa_{\mu\nu}=\partial K_{\mu}/\partial\xi_{\perp\,\nu}$ with the derivatives evaluated for ${\bm\xi}_{\perp}={\bf 0}$; the subscripts $\mu,\nu=1,\ldots,N-1$ enumerate the components of $\bm\xi_{\perp}$ and the transverse components of ${\bf K}$ in the co-moving frame.

The case where the observation point $\qb$ lies in the basin of attraction of the initially occupied state $A_1$ is more complicated. We seek the solution of Eqs.~(\ref{eq:FPE}) for small $D$ in the eikonal form,
$ \rho(\qb,t|\qb_0,t_0)=\exp[-S_{\rm F}(\qb)/D]$,  $\rho(\qb_f,t_f|\qb,t)=\exp[S_{\rm B}(\qb)/D]\rho_2(\qb_f)$,
with $S_{\rm F,\,B}= S_{\rm F, \,B}^{(0)} + D S_{\rm F, \,B}^{(1)} +\ldots$. To the lowest order in $D$ we have
\begin{eqnarray}
\label{eq:Hamilton-Jacobi}
 H\left(\qb,\,\partial_{\qb}S_{{\rm F},\,{\rm B}}^{(0)}\right)=0,\qquad
 H(\qb,\pb)= \pb^2 + \pb{\bf K}(\qb).
\end{eqnarray}
Equation (\ref{eq:Hamilton-Jacobi}) has a form of a Hamilton-Jacobi equation. The function $S_{\rm F}^{(0)}(\qb)$ is the least action for reaching $\qb$ starting from $\qb_{A_1}$ \cite{Freidlin_book}; $S_{{\rm B}}^{(0)}(\qb)$ is the least action for reaching $\qb_S$ from $\qb$. The MPSP is given by the heteroclinic Hamiltonian trajectory from $\qb_{A_1}$ to $\qb_{\cal S}$. Clearly, on the MPSP $S_{{\rm B}}^{(0)}(\qb)-S_{\rm F}^{(0)}(\qb)$ is maximal, i.e.,
$ \partial_{\qb}S_{\rm F}^{(0)} = \partial_{\qb}S_{\rm B}^{(0)}$.
From Eq.~(\ref{eq:Hamilton-Jacobi}), the MPSP direction $\hat\xib_{\parallel}$ and the velocity  are given by the expression
$ 2\partial_{\qb}S_{\rm F, \, B}^{(0)}(\qb) +{\bf K}(\qb) = v(\xi_{\parallel})\hat\xib_{\parallel}$.

Matrix $\widehat Q$ in Eq.~(\ref{eq:2}) can be found from Eq.~(\ref{eq:Hamilton-Jacobi}),
$ \widehat{Q} = \widehat{Q}_{\rm F}-\widehat{Q}_{\rm B}, \quad (Q_{\rm F,\,B})_{\mu\nu}=D^{-1}\partial^2 S_{\rm F,\,B}^{(0)}/\partial\xi_{\perp\mu}\partial\xi_{\perp\nu} $
with the derivatives calculated on the MPSP, i.e., for $\xib_{\perp} = {\bf 0}$. The prefactor in $p_{12}(\qb,t)$ is determined by $S_{\rm F}^{(1)}-S_{\rm B}^{(1)}$. Central for obtaining it is the relation
$ v\partial_{\xi_{\parallel}}\widehat{Q} + 2D\left(\widehat{Q}_{\rm F}^2-\widehat{Q}_{\rm B}^2\right)+ \hat\kappa^{\dagger}\widehat{Q} + \widehat{Q}\hat\kappa=0$
that follows from Eq.~(\ref{eq:Hamilton-Jacobi}).

An important property of the distribution (\ref{eq:2}) is probability current conservation, $\int d\xib_{\perp}v(\xi_{\parallel})\p12 =W_{12}$  independent of the position on the path. This result could be expected from the physical picture of switching. Because the velocity  $v(\xi_\|)$ decreases near the stationary states  $\qb_{A_{1,2}}$,  $\qb_{S}$, the distribution increases there. Close to the saddle point Eq.~(\ref{eq:2}) should be modified.  For
systems with detailed balance one should replace $v$  by $\zeta l_D /t_r$  with $\zeta \sim 1$.

The switching probability distribution (\ref{eq:1}) is measured for a high-Q micro-electro-mechanical torsional oscillator (Q = 9966) driven into parametric resonance \cite{Chan2007}. As shown in Fig.~\ref{fig:1}c, the device consists of a 3.5-$\mu m$ thick, 200-$\mu m^2$  heavily doped polysilicon plate suspended by two thin torsional rods on opposite edges. In addition to the restoring torque of the torsional springs, the top plate is also subjected to an electrostatic torque when a voltage is applied to the left electrode in Fig.~\ref{fig:1}d. The voltage is modulated at a frequency $\omega / 2 \pi$  (44346.800 Hz) close to twice the natural oscillation frequency (22178.578 Hz) of the plate. The orientation angle of the plate $\theta(t)$ is detected capacitively through the right electrode in Fig.~\ref{fig:1}d.

When the modulation is sufficiently strong, the plate oscillates at half the modulation frequency as a result of parametric resonance. Since the modulation is invariant upon a shift in time by its period, there exist two stable oscillation states that have the same amplitude but differ in phase by  $\pi$. The dynamics is well described by the rotating wave approximation \cite{Chan2007}. It is characterized by two dynamical variables $X$ and $Y$ (the quadratures), with $\theta(t)=X(t)\cos(\omega t/2) + Y(t)\sin(\omega t/2)$. A lockin amplifier is used to record X and Y every 2 ms. In Fig.~\ref{fig:1}b, we show the two stable oscillation states,  $A_1$ and  $A_2$. They are located symmetrically about the origin in the (X,Y)-plane. The measurement uncertainty is $\sim 80$~$\mu$rad, about $0.6 \%$ of the full scale in Fig. \ref{fig:1}b. All measurements were performed at 77 K and $< 10^{-6}$ torr.

When white noise is added to the excitation voltage, the system can occasionally overcome the activation barrier and switch from one stable state to the other. The noise intensity is chosen to ensure that the mean residence time in each state ($\sim 10$~s) is much larger than the relaxation time ($\sim 1$~s) of the system. The oscillator dynamics in the rotating frame is described by Eq.~(\ref{eq:Langevin}) with $q_1=CX, q_2=CY$ and ${\bf K}=-\zeta^{-1}\qb +\hat\epsilon\n g$, where $g=q^4/4-(1-\mu)q_1^2/2+(1+\mu)q_2^2/2$ and $\hat\epsilon$ is the permutation tensor; from the harmonic and parametric resonances of the oscillator we found $C=176.349, \zeta =4.968, \mu=0.9367$.

Transitions are identified when the oscillator begins in the vicinity of the stable state $A_1$ (within the left green circle in Fig.~\ref{fig:1}b) and subsequently arrives at state   $A_2$ (within the right green circle). Figure~\ref{fig:1}a shows the switching probability distribution derived from more than 6500 transitions. While in each transition the system follows a different trajectory, the trajectories clearly lie within a narrow tube. The maximum of the distribution gives the MPSP. In Fig.~\ref{fig:1}b, the location of this maximum is plotted on top of the MPSP obtained from theory. The MPSP emerges clockwise from $A_1$  and spirals toward the saddle point at the origin. Upon exit from the saddle point, it makes an angle and continues to spiral clockwise toward  $A_2$. There is excellent agreement between the measured peak and the MPSP obtained from theory, with no adjustable parameters.

Close to the stable states the peaks of the distribution at successive turns of the MPSP overlap, preventing the accurate determination of the MPSP. The plot in Fig. \ref{fig:1}a,b has excluded the portions of trajectories prior to escaping from the initial state   $A_1$ and upon arriving at the final state   $A_2$, which are bound by the two blue lines. Such cutoff also eliminates the large peaks of the distribution centered at  $A_1$  and  $A_2$, which arise because the oscillator spends most of its time fluctuating about  $A_1$  and   $A_2$. These peaks are not relevant to switching dynamics.

\begin{figure}[h]
\includegraphics[width=3.2in]{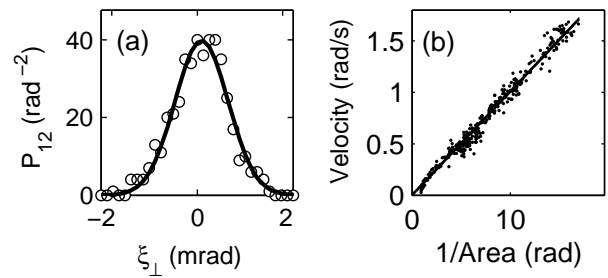}
\caption{\label{fig:2} (a) The cross-section of the switching probability distribution along the purple line in Fig.~\ref{fig:1}b transverse to the MPSP. The solid line is a Gaussian fit. (b) Velocity on the MPSP vs. inverse area under cross-sections of the switching probability distribution. The solid line is a linear fit forced through the origin. }
\end{figure}

The switching probability distribution in our multi-variable system displays important generic features. Figure~\ref{fig:2}a shows the distribution cross-section along the purple line transverse to the MPSP in Fig.~\ref{fig:1}b.  It is well-fitted by a Gaussian. Gaussian distributions with different height and area are observed also in other cross-sections except close to the saddle point. Figure~\ref{fig:2}b plots the area under the Gaussian distribution versus the reciprocal measured velocity on the MPSP, for different cross-sections. The linear dependence agrees with Eq.~(\ref{eq:2}) and indicates that the probability current from the initially occupied attractor to the empty one is constant. Near the saddle point the velocity is significantly diminished and the motion becomes diffusive, leading to strong broadening and increase in height of the distribution.

We find that the probability current concentrates within a narrow tube deep into the basins of attraction of $A_1$ and   $A_2$. In the basin of attraction to $A_2$ but not too close to $A_2$, much of the probability distribution carries the switching current. However, the probability distribution deep inside the basin of attraction of $A_1$ is largely associated with fluctuations about $A_1$ that do not lead to switching. The part of the distribution responsible for the switching current is an exponentially small fraction of the total distribution. Nevertheless our formulation makes it possible to single out and directly observe this fraction.

\begin{figure}[h]
\includegraphics[width=2.6in]{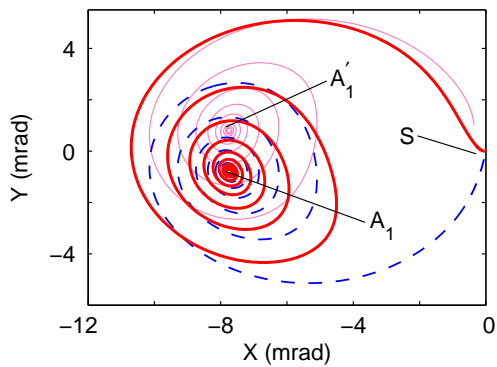}
\caption{\label{fig:3} (Color online) Comparison of the MPSP, the time-reversed path and the deterministic downhill path. The section of the most probable switching path from $A_1$ to S is shown as a thick solid  line. Upon changing the sign of dissipation, the attractor is shifted to a new location  $A_1^{\prime}$ and becomes an unstable state. The fluctuation-free path with reversed dissipation from  $A_1^{\prime}$ to S is shown as the thin solid line. The dashed line represents the deterministic downhill path from S to $A_1$.  }
\end{figure}

Another generic feature of the observed distribution is characteristic of systems far from thermal equilibrium. For equilibrium systems, the most probable fluctuational path ``uphill'' from an attractor to the saddle point is the time reversal of the fluctuation-free ``downhill'' path from the saddle point back to the attractor. More precisely, it corresponds to the change of the sign of dissipation term in the equation of motion, cf. Ref.~\onlinecite{Onsager1953}. Our parametric oscillator is far from thermal equilibrium and lacks detailed balance. Upon sign reversal of dissipation, even the attractors are shifted away from the original locations (from $A_1$ to $A_1^{\prime}$  in Fig.  \ref{fig:3}). Our data show that the uphill section of the MPSP is distinct from both the dissipation-reversed path and the downhill noise-free path from the saddle to the stable state.

The observation of the MPSP reported here provides, in some respects, an experimental basis for the broadly used concept of a reaction coordinate, which can be associated with the coordinate along the MPSP. An advantageous feature of our approach is that it does not rely on a specific model of the fluctuating system. The only characteristics used are the positions of the stable states in phase space, which are usually accessible to direct measurement. The approach applies to systems far from thermal equilibrium as well as to equilibrium systems. Measuring the switching trajectories can help to determine the model globally, far from the stable states. It can also provide an efficient way of controlling the switching rates by affecting the system locally on the most probable switching path.

In conclusion, we propose a theory and report the first observation of the distribution of paths followed by a system in switching between stable states. We show that, in phase space, switching paths form a narrow tube centered at the MPSP. The observed features of the paths distribution and the position of the MPSP for a well-characterized system are in full agreement with the theory.

This research was supported in part by NSF DMR-0645448 (HBC) and NSF PHY-0555346 (MID).


\end{document}